\documentclass[fleqn,10pt]{wlscrep}

\usepackage[utf8]{inputenc}
\usepackage[T1]{fontenc}
\usepackage{multirow}
\usepackage{float}
\usepackage{booktabs}
\usepackage{makecell}
\usepackage{hhline}
\usepackage{placeins}
\usepackage{soul}
\usepackage{xcolor}
\usepackage[numbers]{natbib}  
\DeclareUnicodeCharacter{2212}{-}

\DeclareUnicodeCharacter{FF0C}{,}

\title{Estimating the effect of lymphovascular invasion on 2-year survival probability under endogeneity: a recursive copula-based approach}

\usepackage{authblk}

\author[1,+]{Yang Ou}
\author[2]{Lan Xue}
\author[1]{Carmen Tekwe}
\author[1]{Kedir N Turi}
\author[1,+,*]{Roger S. Zoh}

\affil[1]{Indiana University, Department of Epidemiology and Biostatistics, Bloomington, IN 47405, USA}
\affil[2]{Oregon State University, Department of Statistics, Corvallis, OR 97331, USA}
\affil[+]{These authors contributed equally to this work}
\affil[*]{Corresponding author: rszoh@iu.edu}


\begin{abstract}
Lymphovascular invasion (LVI) is an important prognostic marker for head and neck squamous cell carcinoma (HNSC), but the true effect of LVI on survival may be distorted by endogeneity from unmeasured confounding. Conventional one-stage, conditional models and instrument-based two-stage estimators are prone to bias under endogeneity, and sufficiently strong instruments are often unavailable in practice. To address these challenges, we propose a semiparametric recursive copula framework that jointly specifies marginal models for both LVI (an endogenous “treatment”) and a binary 2-year survival outcome and links them via a flexible copula to correct for latent confounding and accommodate censoring without requiring strong instruments. In two simulation studies, we systematically varied sample sizes, censoring rates (0\%–60\%), and endogeneity strengths and assessed robustness to moderate model misspecification. The copula framework resulted in less bias and better interval coverage than both the one- and two-stage approach while maintaining robustness under moderate model misspecification. We applied the copula framework to estimate the effects of LVI on 2-year survival in HNSC cases with associated clinical and microRNA data from The Cancer Genome Atlas (n=215), finding that LVI significantly reduces 2-year survival by 47\% (95\% confidence interval: -0.61 to -0.29). The estimated positive dependence parameter suggests that the attenuation is driven by residual dependence between unobserved components of LVI and survival. Overall, the copula framework for survival outcomes provides more credible effect estimates than other methods in the absence of strong instruments, mitigating biases due to endogeneity and censoring, and strengthening quantitative evidence for HNSC research.
\end{abstract}

\usepackage[utf8]{inputenc}
\begin{document}
\flushbottom
\maketitle
%
%
\thispagestyle{empty}
\noindent 

\section*{Introduction}

Head and neck squamous cell carcinoma (HNSC) negatively impacts quality of life due to the formation of malignancies on the mucosal surfaces of critical anatomical structures involved in speech, breathing, and swallowing  \cite{johnson2020head, babin2008quality}. HNSC is the seventh most common cancer worldwide, with a 5-year survival rate below 50\%, substantially lower than the average 5-year survival rate of 68\% for all cancers  \cite{barsouk2023epidemiology, shibata2021immunotherapy, siegel2023cancer}. Lymphovascular invasion (LVI), in which tumor cells penetrate lymphatic or vascular systems, is recognized as an early stage of metastatic progression and serves as a prognostic marker of HNSC that correlates with poorer survival outcomes  \cite{pisani2020metastatic, karahatay2007clinical, zhang2020key}.

Understanding how LVI impacts patient survival is critical for accurate prognostication, which directly influences treatment decisions and patient outcomes \cite{ha2004role}. However, LVI is not merely a binary pathological finding; it is a manifestation of an underlying invasive tumor phenotype \cite{zhang2020key, adel2015evaluation}. The same latent biological aggressiveness (e.g., enhanced invasion/migration capacity, lymphangiogenic signaling, immune evasion, and other microenvironmental or molecular processes) can simultaneously increase the likelihood of observing LVI and worsen patient survival \cite{pectasides2014markers,jumaniyazova2023role,tsai2025immune, cai2025advancing}. Because these underlying biological processes are only partially observed in routinely collected clinical covariates \cite{tsai2025immune}, standard survival models that treat LVI as exogenous may be vulnerable to residual confounding or endogeneity, reflected in residual dependence between the LVI process and the survival process after adjustment \cite{fewell2007impact}. In medical research, especially oncology research, unmeasured confounders can simultaneously affect disease markers and survival outcomes due to complex biological interdependencies, introducing endogeneity \cite{huang2023prognostic, huang2021impact, cho2021endogenous}. In statistical analyses, endogeneity occurs when explanatory variables correlate with the error terms of a regression model, typically due to omitted variables, measurement errors, or feedback mechanisms between exposure and outcome \cite{terza2008two}. Conceptually, an unobserved aggressive tumor state can act as a shared cause of both LVI and survival, inducing correlated unobserved components (or dependent errors) between the LVI and survival equations \cite{terza2008two}. Clinically, LVI is often observed alongside other adverse pathologic features, which is consistent with the presence of a shared underlying tumor aggressiveness that also drives prognosis \cite{ting2021perineural,prasad2023trends}. Translating well-documented clinical findings into unbiased prognostic estimates requires careful handling of LVI’s potential endogeneity in survival models.

Existing studies of LVI often use classical regression frameworks and treat LVI as an exogenous predictor \cite{huang2021impact}. These models assume the absence of endogeneity, effectively treating LVI as exogenous and potentially overlooking its association with underlying biological determinants or latent confounders that jointly influence LVI and survival \cite{ha2004role}. In survival analyses with a binary covariate, such as LVI, assuming the absence of endogeneity also implies that the error term is independent of the covariate; however, if LVI and survival share unobserved common causes, their corresponding error terms may become dependent and correlated, violating independence \cite{lin1998assessing}. As a result, treating LVI as exogenous can misattribute the prognostic impact of unmeasured aggressiveness to LVI itself, yielding biased and potentially misleading estimates of LVI’s survival effect \cite{terza2008two}. Therefore, failing to account for endogeneity can yield biased, potentially misleading estimates of treatment effects, which are particularly problematic in the context of survival analysis, as treatment planning relies heavily on robust prognostic models \cite{terza2008two}.

To address endogeneity in biomedical research, several methodological strategies have been employed, including instrumental variable (IV) approaches, propensity score adjustments, and semiparametric or machine learning techniques, with sensitivity analyses used primarily for robustness assessment \cite{becker2022revisiting, yan2024pmdags, karmakar2021identification}. Two-stage estimation methods based on IVs introduce external variables that do not directly affect the outcome, modeling endogenous variables in the first stage to reduce bias due to unobserved confounders. However, valid IVs must satisfy strict assumptions, including relevance (i.e., being strongly associated with the endogenous predictor), independence/exogeneity (i.e., not sharing unmeasured causes with the outcome), and the exclusion restriction (i.e., affecting the outcome only through the endogenous variable) \cite{baiocchi2014tutorial,martens2006instrumental,brookhart2010instrumental}.
Consequently, these assumptions can be difficult to justify empirically in many biomedical settings, making credible instruments challenging to identify \cite{martens2006instrumental,brookhart2010instrumental}. Even when valid IVs can be identified, two-stage methods face substantial limitations. Traditional two-stage techniques are primarily oriented toward continuous outcome variables, and the commonly used two-stage least squares method (2SLS) assumes a linear relationship between variables\cite{terza2008two}. Therefore, without substantial adjustments, these methods are often inappropriate for nonlinear or non-continuous outcomes, including hazard rates in survival analysis. In addition, standard two-stage substitution estimators such as 2SLS or 2SPS effectively treat the first-stage fitted values as if they were observed without error in the second stage and rely on strong assumptions about the error structure and correct specification of both stages, assumptions that are generally not satisfied in nonlinear or survival settings, leading to bias and loss of robustness \cite{terza2008two,chapman2016treatment,cai2011two}. In practical applications, these strict prerequisites are often difficult to fully meet, weakening the robustness of the method and the effectiveness of the results. These challenges motivate joint modeling approaches that explicitly allow correlated unobserved components between the LVI and survival processes, while also accommodating potentially nonlinear covariate effects via flexible semiparametric specifications \cite{radice2016copula}.

We leveraged the increasing availability of large-scale transcriptomics datasets, such as The Cancer Genome Atlas (TCGA), and applied modern statistical tools to revisit the role of LVI in HNSC survival. Although microRNAs (miRNAs) have been shown to modulate tumor aggressiveness and are associated with both LVI and survival outcomes \cite{karmakar2021identification, Beaubier2019}, previous studies have employed fragmented analytical strategies that failed to account for the joint distribution of LVI and survival or for the effects of confounding molecular signals \cite{sabatini2020human,ha2004role}. 
In this study, we applied and evaluate a semiparametric recursive copula framework that jointly models a binary LVI process and a survival process, allowing residual dependence between their unobserved components and accommodating nonlinear covariate effects via flexible functions. 
We first report the findings of our simulation experiments, assessing bias and efficiency under various censoring and dependence scenarios.
In the TCGA-HNSC application, we estimate a substantial negative effect of LVI on 2-year survival (approximately a 47\% decrease in survival probability) after accounting for endogeneity and incorporating miRNA and other clinical covariates.
Our findings highlight the importance of modeling strategies that capture structural dependencies and nonlinear biomarker effects, providing new insights with clinical implications for risk stratification and treatment planning in HNSC.

\section*{Results}

\subsection*{Simulation Study 1}

\begin{figure}[H]
\includegraphics[width=\textwidth]{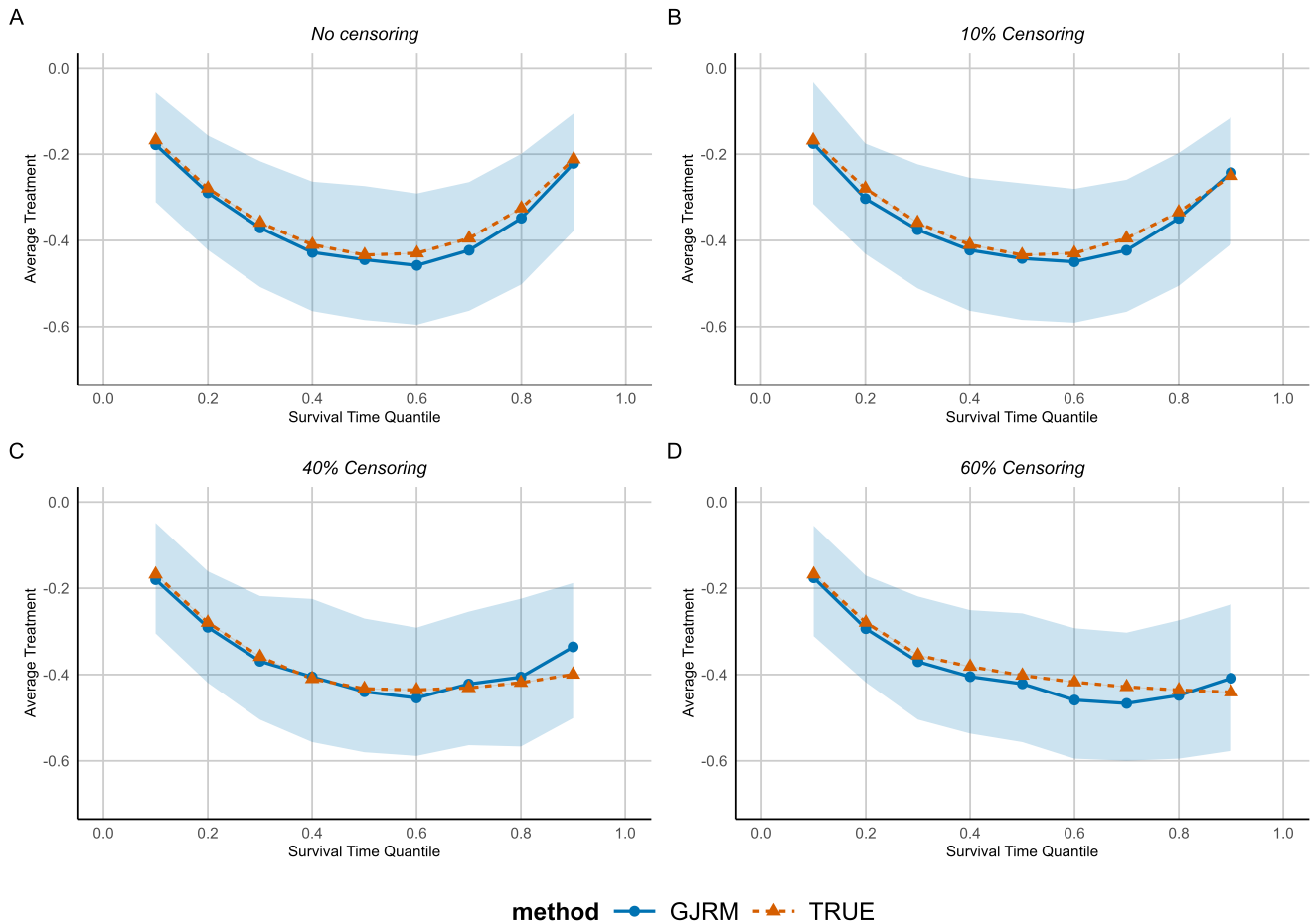}
\caption{\textit{SATE estimates across various censoring rates (All Coefficients Non-Zero). A) No censoring. B) 10\% censoring. C) 40\% censoring. D) 60\% censoring. Datasets are simulated from a bivariate, Gaussian, copula, probit–probit marginal link function, assuming $\rho$=0.5, and all covariates have non-zero effects. For all simulations, the sample size is n=200 for 200 independent replicates. The x-axis denotes survival time quantiles (0.1, $...$, 0.9) based on the observed survival time. Each value corresponds to a cutoff point used to define binary survival outcomes (e.g., alive vs. dead) at that specific time threshold. Orange triangles represent the true average treatment effect. Blue circles represent the values obtained using the generalized joint regression model (GJRM). SATE, sample average treatment effect.}
}
\label{fig30000all200}
\end{figure}

\begin{figure}[H]
\includegraphics[width=\textwidth]{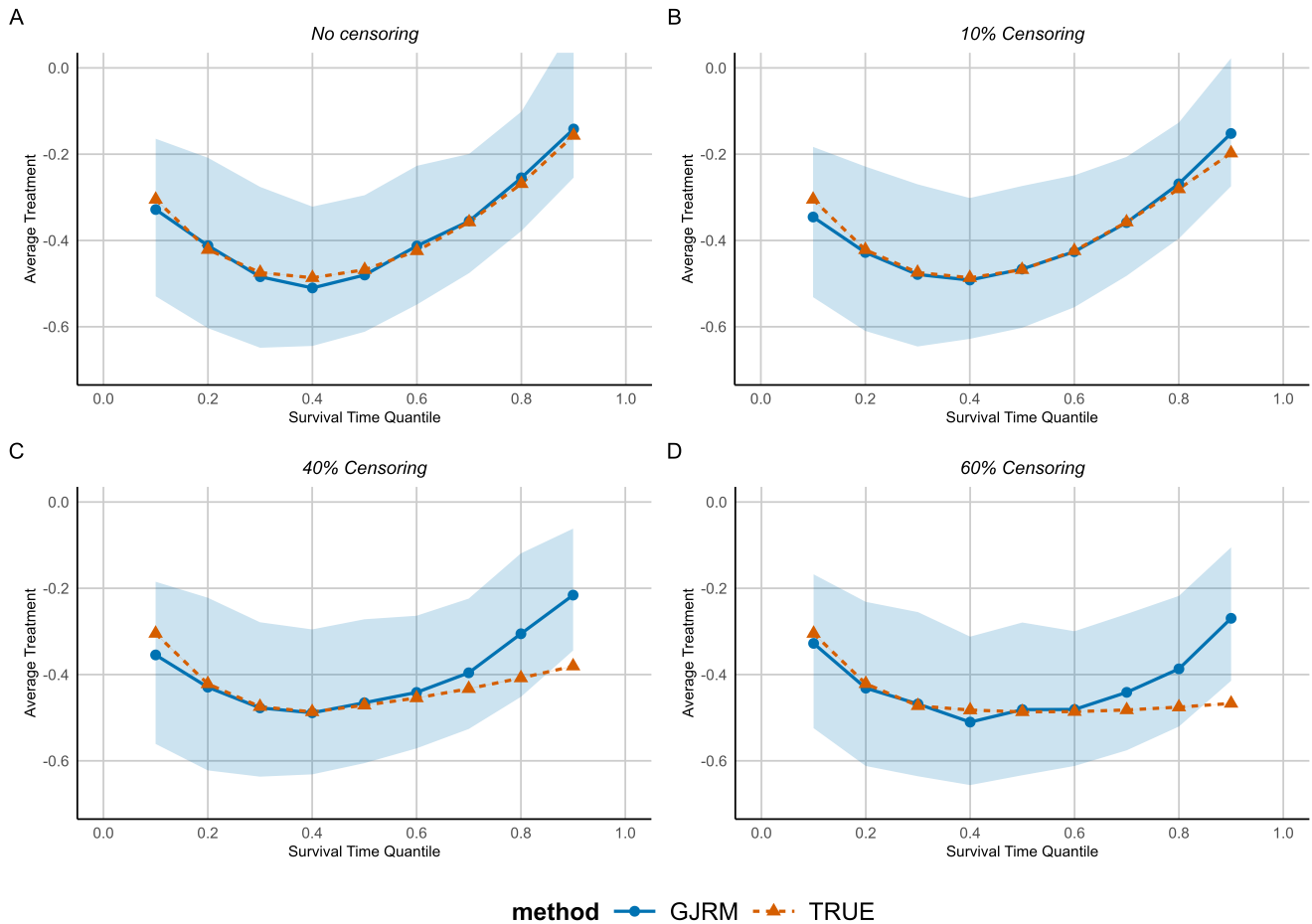}
\caption{\textit{SATE estimates across various censoring rates (Half Coefficients Zero). A) No censoring. B) 10\% censoring. C) 40\% censoring. D) 60\% censoring. Datasets are simulated from a bivariate, Gaussian, copula, probit–probit marginal link function, assuming $\rho$=0.5, and all covariates have non-zero effects. For all simulations, the sample size is n=200 for 200 independent replicates. The x-axis denotes survival time quantiles (0.1, $...$,0.9) based on the observed survival time. Each value corresponds to a cutoff point used to define binary survival outcomes (e.g., alive vs. dead) at that specific time threshold. Orange triangles represent the true average treatment effect.  Blue circles represent the values obtained using the generalized joint regression model (GJRM). SATE, sample average treatment effect.}}

\label{fig30000half200}
\end{figure}

In the first simulation study, we evaluate the ability of the semiparametric copula approach to estimate the sample average treatment effect (SATE) under different censoring rates. To assess robustness, we consider two scenarios: (1) a fully specified model, in which all covariates have non-zero effects; and (2) a sparse, misspecified model, in which only half of the covariates are truly informative, and the remaining covariates have zero coefficients. In both settings, the copula method was applied to datasets with a sample size of 200, and estimates were averaged over 200 replicates. We examined various censoring rates, including 0\% (no censoring), 10\%, 40\%, and 60\%, and plotted the corresponding SATE estimate curves and their pointwise 95\% confidence intervals (CIs) across the survival time quantile interval (0, 1). All simulation results are evaluated against the true average treatment effect curve, which serves as the reference in all figures. Figure \ref{fig30000all200} shows the results for the fully specified model, Figure \ref{fig30000half200} shows the results for the partially specified model, and additional results can be found in Supplementary Figures S3 and S4.

When the censoring rate is low (e.g., 0.1\% or 10\%), the copula method yields estimates that closely align with the benchmark curve, with minimal error. Estimation errors become more pronounced as the censoring rate increases, such as at censoring rates in the upper tail (e.g., 60\%), which reflect right censoring that limits information. However, the copula method continues to produce reasonably accurate estimates in the central portion of the survival distribution, even under high censoring rates (e.g., 50\% or 60\%), and the overall U-shaped trend of the SATE curve is preserved across all scenarios, highlighting the method’s robustness in the presence of moderate to high levels of censoring. The benchmark CIs are extremely narrow and are, therefore, barely visible in the figures.

Taken together, the similarity of results across different benchmark sizes and model specifications highlights the robustness and stability of the copula-based approach. The method performs well under moderate censoring, and even under high censoring rates, it continues to preserve the overall trend, underscoring its practical value in survival analysis.

\subsubsection*{Simulation Study 2}

In the second simulation study, we compare the performance of three methods: the naïve one-stage method, a two-stage predictor substitution (2SPS), and the copula-based joint model. The 2SPS we consider is the natural analogue of 2SLS for a binary outcome, where the treatment is first regressed on the instruments and the fitted values are then entered into a logistic regression for the binary outcome. These two-stage IV estimators, including 2SLS and 2SPS, are intended for settings with strong external instruments \cite{terza2008two,martens2006instrumental}. In our simulation design, we intentionally assume the absence of strong instruments, thereby violating the key identification assumptions underlying two-stage methods. As a result of this violation, the 2SPS approach produces highly biased and unstable estimates across all scenarios, consistent with the theoretical limitations of this approach in the absence of strong instruments \cite{li2022binary}. Given these findings, we report the results of the 2SPS approach in Supplementary Tables S18–S20, as a reference only, focusing instead on the comparison between the naïve one-stage method and the copula-based approach, which are more representative of the application settings we aim to address.

We systematically evaluate the accuracy of each method for estimating the coefficient of the dependent variable across varying values of the copula correlation parameter, denoted by $\rho$. Specifically, we set $\rho$ = 0, 0.5, and 0.7 and compare the performance of the one-stage, two-stage, and copula methods for estimating the $Y_1$ coefficient. The primary evaluation metric is the signed bias of the estimated coefficient. The simulation results consistently show that the copula method outperforms the single-stage approach, especially under moderate to high censoring and stronger dependency ($\rho$ = 0.7). Table \ref{tab:differror_0} presents the estimation bias results for no endogeneity ($\rho$ = 0), and Table \ref{tab:differror_0.7} presents the estimation bias results for high endogeneity ($\rho$ = 0.7), under different sample sizes (n = 200, 500, 1000), censoring rates $(0\%, ~30\%, ~60\%)$, and survival time quantiles $(0.25, ~0.5,~0.75)$. The results for moderate endogeneity ($\rho$ = 0.5) are reported in Supplementary Table S19.

\begin{table}[H]
\centering
\caption{
Estimated signed bias relative to the true simulated coefficient of $Y_1$ under no endogeneity ($\rho=0$).}

\label{tab:differror_0}
\begin{tabular}{cccccccc}
\toprule
\multirow{2}{*}{$n$} & \multirow{2}{*}{Censoring Rate} 
& \multicolumn{2}{c}{Survival Cut-off Quantile = 0.25} 
& \multicolumn{2}{c}{Survival Cut-off Quantile = 0.50} 
& \multicolumn{2}{c}{Survival Cut-off Quantile = 0.75} \\
\cmidrule(lr){3-4} \cmidrule(lr){5-6} \cmidrule(lr){7-8}
& & One Stage & Copula & One Stage & Copula & One Stage & Copula \\
\midrule
200  & 0\%  & -0.0222 & -0.1273 & -0.0125 & -0.1030 & -0.0206 & -0.0523 \\
     & 30\% & -0.1461 & -0.1140 & -0.1380 & -0.0951 & -0.0436 & -0.1397 \\
     & 60\% & -0.3939 & -0.1209 & -0.3344 & -0.1068 & -0.1516 & -0.1512 \\
500  & 0\%  &  0.0191 & -0.0518 &  0.0137 & -0.0489 &  0.0257 & -0.0262 \\
     & 30\% & -0.0925 & -0.0496 & -0.0986 & -0.0399 & -0.0065 & -0.0331 \\
     & 60\% & -0.3640 & -0.0931 & -0.2939 & -0.0509 & -0.1299 & -0.0931 \\
1000 & 0\%  &  0.0362 & -0.0076 &  0.0253 & -0.0090 &  0.0372 & -0.0393 \\
     & 30\% & -0.0968 & -0.0225 & -0.0746 & -0.0234 & -0.0047 & -0.0219 \\
     & 60\% & -0.3366 & -0.0122 & -0.2805 & -0.0368 & -0.1163 & -0.0670 \\
\bottomrule
\end{tabular}

\vspace{1ex}
\raggedright
\footnotesize
\textit{Results are presented for one-stage and copula methods across different sample sizes $(n = 200, 500, 1000)$, censoring rates $(0\%, 30\%, 60\%)$, and survival time quantiles $(0.25, 0.50, 0.75)$. Because the data were generated under an accelerated failure time model but analyzed with a Cox proportional hazards model, the estimated coefficients were multiplied by $-1$ before bias calculation to ensure a consistent direction.}
\end{table}

\begin{table}[H]
\centering
\caption{
Estimated signed bias relative to the true simulated coefficient of $Y_1$ under high endogeneity ($\rho=0.7$).} 

\label{tab:differror_0.7}
\begin{tabular}{cccccccc}
\toprule
\multirow{2}{*}{$n$} & \multirow{2}{*}{Censoring Rate} 
& \multicolumn{2}{c}{Survival Cut-off Quantile = 0.25} 
& \multicolumn{2}{c}{Survival Cut-off Quantile = 0.50} 
& \multicolumn{2}{c}{Survival Cut-off Quantile = 0.75} \\
\cmidrule(lr){3-4} \cmidrule(lr){5-6} \cmidrule(lr){7-8}
& & One Stage & Copula & One Stage & Copula & One Stage & Copula \\
\midrule
200  & 0\%  & 0.7263 & -0.1209 & 0.7140 & -0.1324 & 0.7276 & -0.1471 \\
     & 30\% & 0.6575 & -0.1434 & 0.6583 & -0.1480 & 0.6993 & -0.1279 \\
     & 60\% & 0.5466 & -0.1253 & 0.5811 & -0.1401 & 0.6541 & -0.1216 \\
500  & 0\%  & 0.7369 & -0.0598 & 0.7448 & -0.0625 & 0.7410 & -0.0831 \\
     & 30\% & 0.6723 & -0.0883 & 0.6716 & -0.0536 & 0.7094 & -0.0407 \\
     & 60\% & 0.5597 & -0.0706 & 0.5939 & -0.0797 & 0.6628 & -0.0673 \\
1000 & 0\%  & 0.7545 & -0.0336 & 0.7515 & -0.0389 & 0.7532 & -0.0240 \\
     & 30\% & 0.6764 & -0.0352 & 0.6734 & -0.0378 & 0.7186 & -0.0149 \\
     & 60\% & 0.5675 & -0.0325 & 0.5969 & -0.0425 & 0.6801 & -0.0308 \\
\bottomrule
\end{tabular}

\vspace{1ex}
\raggedright
\footnotesize
\textit{Results are presented for one-stage and copula methods across different sample sizes (n = 200, 500, 1000), censoring rates (0\%, 30\%, 60\%), and survival time quantiles (0.25, 0.50, 0.75). Because the data were generated under an accelerated failure time model but analyzed with a Cox proportional hazards model, the estimated coefficients were multiplied by -1 before bias calculation to ensure a consistent direction.} 
\FloatBarrier
\end{table}

Under the condition of no dependence or endogeneity ($\rho=0$; Table \ref{tab:differror_0}), the copula method consistently exhibits the lowest estimated bias across all sample sizes and censoring rates, indicating superior estimation efficiency and consistency in this setting. Although the one-stage method performs reasonably well under conditions of minimal censoring, bias increases substantially for this method with higher censoring rates, particularly in smaller samples. Therefore, when $\rho=0$ = 0, the copula-based method generally performs better than the one-stage method.

Under moderate dependence ($\rho=0.5$, Supplementary Table S19), the one-stage approach performs poorly. The overall deviation is between 0.3 and 0.55, and the error is significantly higher than that of the copula method. Although the copula method exhibits a slight increase in bias, it remains within a low range (mostly below 0.2) and decreases further as the sample size increases.

When the error-term correlation is high ($\rho=0.7$, Table \ref{tab:differror_0.7}), the bias of the one-stage method increases substantially, often exceeding 0.7, reflecting limited robustness under conditions of strong dependency or endogeneity. Although the copula method is also affected by high endogeneity, bias remains substantially lower for the copula method than for the one-stage method, and deviation decreases significantly with increasing sample sizes, with bias approaching 0.03 when n = 1000. These results demonstrate the robustness of the copula method and its ability to handle high-dependency scenarios.

\subsection*{Real Data Results}

Given the positive performance of our copula-based approach in simulations, we apply the proposed copula-based method to the analysis of the TCGA-HNSC cohort. Survival data were converted to a binary format to align with the modeling requirements of the generalized joint regression modeling (GJRM) framework, enabling joint analysis of the binary LVI endpoint (treatment) and survival outcomes. Although the median follow-up time in the TCGA-HNSC dataset is approximately 21.2 months \cite{liu2018integrated}, we selected 24 months as the cutoff point due to the widespread use of this cutoff point for defining short-term survival benchmarks in oncology \cite{chen2017treatment, witek2017outcomes}. Using this cutoff point also enables easier comparison with existing survival studies and clinical outcome measures, enhancing the clinical interpretability of our findings at the cost of introducing some missingness. The transformation logic is summarized in Table \ref{table:combined}

\begin{table}[H]
  \centering
  \caption{\bf Classification of 2-Year Survival Outcomes and Descriptive Statistics of Key Variables (N = 215)}
  \label{table:combined}
  \begin{tabular}{@{}llr@{}}
    \toprule
    \multicolumn{3}{l}{\textbf{Panel A. Classification of 2-Year Survival Outcomes}} \\
    \midrule
    \textbf{Description} & \textbf{2-Year Survival Status} & \textbf{Count} \\
    \midrule
    Alive $\geq$ 2 years (Right-censored) & 1  & 157 \\
    Alive $<$ 2 years (Left-censored)     & NA & 126 \\
    Died after 2 years                    & 1  &  55 \\
    Died within 2 years                   & 0  & 147 \\
    \midrule[\heavyrulewidth]
    \multicolumn{3}{l}{\textbf{Panel B. Descriptive Statistics of Variables}} \\
    \midrule
    \textbf{Variable} & \textbf{Definition} & \textbf{Mean (SD)} \\
    \midrule
    LVI                         & =1 if lymphovascular invasion present                      & 0.353 (0.479) \\
    2-Year Survival             & =1 if alive at 2 years                                      & 0.567 (0.497) \\
    Age                         & Age at initial pathologic diagnosis (years)                 & 61.72 (11.88) \\
    Perineural invasion         & =1 if perineural invasion present                           & 0.498 (0.501) \\
    Margin status               & 1 = Positive, 2 = Negative, 3 = Close                       & — \\
    Regional lymph node metastasis        & =1 if regional lymph node metastasis                        & 0.540 (0.499) \\
    \bottomrule
  \end{tabular}

  \vspace{1ex}
  \raggedright
  \footnotesize
  \textit{Notes}: 
  Panel A: Participants were classified at the 24-month cutoff based on survival status and follow-up time. Those alive with follow-up $\geq$ 24 months or who died after 24 months were coded as 1; deaths occurring within 24 months were coded as 0; participants alive with follow-up < 24 months were considered left-censored and treated as missing (NA) \cite{leroux2019organizing}.
Panel B: Data are from The Cancer Genome Atlas Head and Neck Squamous Cell Carcinoma (TCGA-HNSC) cohort (N = 215). Continuous variables are presented as mean (standard deviation); categorical variables are shown as binary indicators.
\end{table}

The model included the following clinical variables: LVI, regional lymph node metastasis, margin status, presence of perineural invasion, and age at initial diagnosis. These variables were selected based on their known clinical relevance for HNSC prognosis. The miRNAs hsa-miR-203a-3p, and hsa-miR-194-5p were included as molecular variables, selected based on preliminary statistical screening results described in the Methods section, previous literature reports, and their known biological functions in HNSC progression \cite{karmakar2021identification, manikandan2016oral, genest2007everything, roncalli2020handbook}.

\subsubsection*{Effect Estimation (SATE)}

\begin{figure}[H]
    \centering    \includegraphics[width=\linewidth]{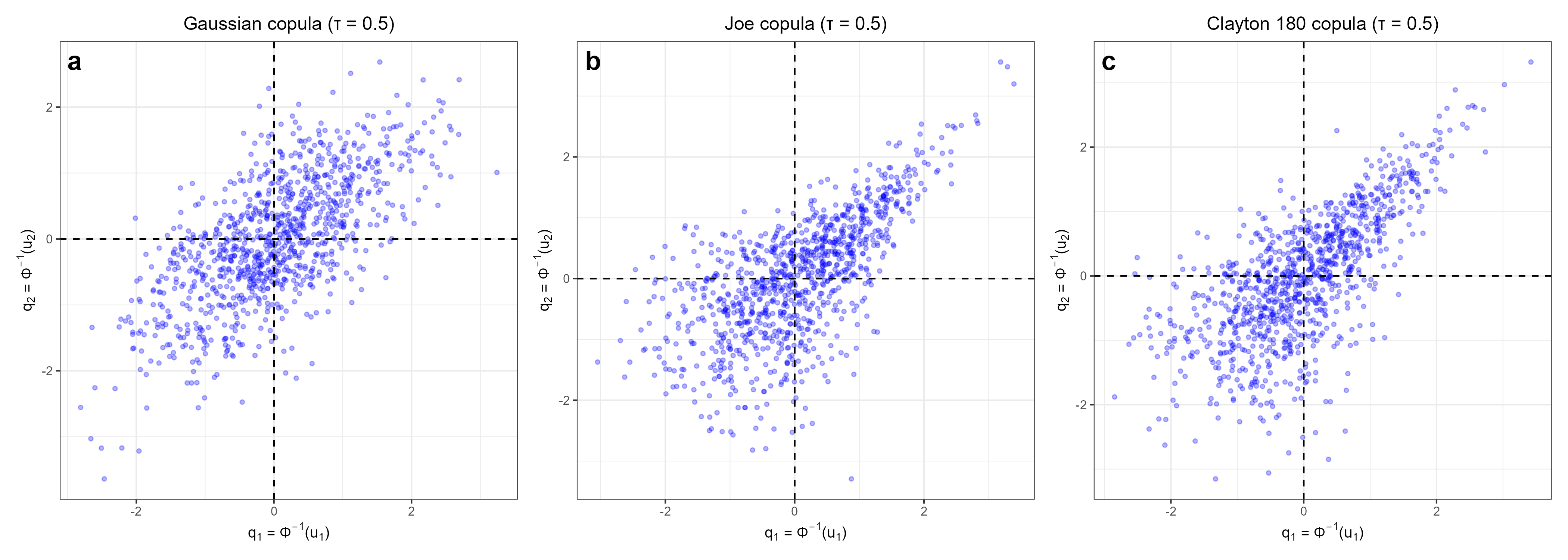}
    \caption{\textit{Scatter plots from simulated copula models at the same Kendall’s $\tau$ $(0.5)$. The copula parameter, $\theta$, differs by family: Gaussian $\rho=0.7071$; Joe  $\theta = 2.8562$; Clayton-180 $\theta=2$. See Supplementary Table S21 for the $\tau-\theta$ relationships.: 
(a) Gaussian copula, (b) Joe copula, and (c) $180^{\circ}$ rotated Clayton copula. 
The Gaussian copula exhibits symmetric dependence, whereas the Joe and rotated Clayton copulas both show pronounced upper-tail clustering but with different patterns and intensities. Axes are labeled as $q_1=\Phi^{-1}(u_1)$ and $q_2=\Phi^{-1}(u_2)$, where $(u_1,u_2)$ are samples from the respective copula families and $\Phi(\cdot)$ is the standard normal cumulative distribution function.}}
    \label{fig:copula_three_panels}
\end{figure}

The results summarized in Table \ref{tablecopula2} present the estimated SATE, Kendall’s $\tau$, and their respective CIs (expressed as CI\%), along with associated goodness-of-fit criteria (Akaike information criteria [AIC] or Bayesian information criteria [BIC]) across various copula models and link function combinations. The models were selected and ranked based on their combined AIC and BIC values, with lower values indicating a better fit. Different link functions reflect distinct assumptions about the latent error distribution in the binary treatment or survival component. Specifically, the logit link assumes a logistic distribution, the probit link assumes a standard normal distribution, and the complementary log-log (cloglog) link implies an asymmetric extreme-value distribution, often used for modeling rare events or hazard-type outcomes. In Table \ref{tablecopula2} we focus on three copula families: Joe and Clayton (180), both suited to capturing upper-tail association (see Figure \ref{fig:copula_three_panels} for tail-concordance diagnostics), and Gaussian, included as a widely used symmetric benchmark. We highlight these three because, under the probit–probit link, the Joe and rotated-Clayton models showed the strongest combined AIC/BIC performance in our data, while Gaussian serves as the canonical symmetric specification for comparison. Although the point estimates of the selected models are similar and their CIs intersect, disparities in estimation results can still be discerned among different copula models. All other copula–link combinations, along with the full grid and diagnostics, are provided in the Supplementary Section S3.

\begin{table}[H]
\centering
\caption{\bf Estimated SATE for 2-Year Survival: Summary Table}
\label{tablecopula2}
\begin{tabular}{llcccc}
\toprule
\textbf{Copula Model} & \textbf{Link} & \textbf{SATE (95\% CI)} & \textbf{AIC} & \textbf{BIC} & \textbf{Kendall's \(\tau\) (95\% CI)} \\
\midrule
Joe           & probit-probit & $-0.469\,(-0.595,\,-0.316)$ & 472.88 & 539.32 & $0.484\,(0.272,\,0.738)$ \\
Clayton (180)  & probit–probit  & $-0.474\,(-0.590,\,-0.324)$ & 473.20 & 539.48 & $0.490\,(0.164,\,0.703)$ \\
Gaussian      & probit-probit & $-0.474\,(-0.610,\,-0.293)$ & 478.27 & 542.37 & $0.439\,(0.156,\,0.731)$ \\
Student-t      & probit-logit  & $-0.450\,(-0.604,\,-0.265)$ & 478.29 & 542.21 & $0.400\,(0.017,\,0.700)$ \\
\bottomrule
\end{tabular}

\vspace{1ex}
\raggedright
\footnotesize
\textit{The estimated sample average treatment effect (SATE), Kendall’s $\tau$, Akaike information criteria (AIC), and Bayesian information criteria (BIC) were obtained using various copula models for The Cancer Genome Atlas–Head and Neck Squamous Cell Carcinoma data, with 95\% confidence intervals (CIs) shown in parentheses.  Rows ranked by AIC (ties by BIC; lower = better).  Joe and Clayton (180°) because both capture upper-tail association and had the best AIC/BIC under probit–probit (pp); Gaussian is a common symmetric benchmark. Student-t allows for symmetric but heavier-tailed dependence than the Gaussian.  Other copula–link combinations and diagnostics are in the Supplement Table S11.}
\end{table}

In terms of model selection, for each copula, different combinations of link functions yielded consistent SATE point estimates and CIs, but results differed across copula families. For example, the $\tau$ value of the Clayton 180 model (0.490) was slightly higher than that of the Joe model (0.484). Although these values are numerically close, Kendall’s $\tau$ is a global measure of concordance that is not directly comparable across families, as distinct copulas with the same $\tau$ can exhibit very different dependence structures \cite{genest2007everything, roncalli2020handbook,venter2002tails}. Among the probit-probit models, the Joe copula yielded the lowest AIC (472.88), and the Clayton (180) copula with a probit–probit link produced the second-lowest AIC (473.20). AIC and BIC are information-based criteria that balance model fit and complexity, with lower values indicating a better trade-off between goodness-of-fit and parsimony. AIC tends to emphasize predictive accuracy, whereas BIC imposes a stronger penalty for model complexity. These models also exhibited low BIC values, indicating strong relative fit. Under the probit–probit link, the Joe copula showed moderately strong dependence (Kendall’s $\tau$ equal to 0.484, 95\% CI 0.272–0.738) and estimated the SATE at -0.465 (95\% CI: -0.582 to -0.310), indicating that the 2-year survival rate of patients with LVI is 46.5\% lower than that of patients without LVI. Likewise, the Clayton (180) model with probit–probit link showed comparable strength of dependency ($\tau$ = 0.490, 95\% CI: 0.164–0.703) and provided a SATE estimate of -0.474 (95\% CI: -0.590 to -0.324), indicating that the 2-year survival rate of patients with LVI is 47.4\% lower than that of patients without LVI. The Gaussian copula, utilizing a probit–probit link, is also highlighted, due to its theoretical simplicity and wide applicability in biomedical research \cite{peng2022latent, marra2023flexible}. This model displayed moderate dependency strength ($\tau$ = 0.439, 95\% CI: 0.156–0.731), despite having somewhat higher AIC (478.27) and BIC (542.37) values, and yielded an estimated SATE of -0.474 (95\% CI: -0.610 to -0.293), indicating that the 2-year survival rate of patients with LVI is 47.4\% lower than that of patients without LVI.

Figure \ref{fig:copula_three_panels} illustrates the dependence structures generated by Gaussian, Joe, and Clayton (180) copula models under the same Kendall's $\tau$ (0.5). Kendall's $\tau$ standardizes the overall level of concordance across families, but dependence manifests quite differently in their shapes. The Gaussian copula (Figure \ref{fig:copula_three_panels}a) exhibits symmetric dependence around the diagonal, consistent with a linear correlation and no tail concentration. By contrast, the Joe copula (Figure \ref{fig:copula_three_panels}b) shows a pronounced upper-tail dependence, with observations clustering in the upper-right quadrant. The rotated Clayton copula (Figure \ref{fig:copula_three_panels}c) also displays an upper-tail concentration, but with a different pattern and intensity than the Joe copula. These difference underscores the importance of reporting both $\tau$ values and family-specific tail characteristics when comparing copula models \cite{roncalli2020handbook,venter2002tails}.

Overall, although the estimated SATE values across different copula models showed considerable consistency, subtle differences in dependency measures and model fits were evident. Notably, the Joe and Clayton (180) copulas consistently outperformed other specifications in terms of AIC and BIC, indicating that these models are robust choices for estimating the SATE of LVI on 2-year survival rates. The Gaussian copula remains valuable as a benchmark, due to its interpretability and general use, reinforcing the robustness of the treatment effect estimates across different modeling approaches.

\subsubsection*{Covariate Effects on Treatment and Outcome Models}

We estimated covariate effects using a combination of parametric and nonparametric modeling strategies. Categorical clinical variables, including regional lymph node metastasis, margin status, perineural invasion, and LVI, were modeled parametrically to enable formal statistical inference. By contrast, continuous variables, such as miRNA expression levels and age at first diagnosis, were incorporated nonparametrically using smooth functions to allow for flexible modeling of potentially nonlinear relationships. This semiparametric approach was implemented within a copula-based framework to jointly model the treatment and outcome processes while accounting for complex marginal structures. Copula selection remained an important factor in accurately estimating the SATE \cite{marra2020estimating, marra2021did}.

As shown in Table \ref{table:copula_compare}, the parametric results from the treatment equation were consistent across the Gaussian, Joe, and Clayton 180 copula models. Regional lymph node metastasis (p=0.0011) and perineural invasion (p=0.0031-0.0038) were significantly and positively associated with treatment assignment in all models. Negative surgical margin status was significantly associated with a lower likelihood of receiving treatment (p<0.001), whereas close margins showed a similarly negative but more modest association (p=0.0266-0.0357).

\begin{table}[H]
\centering
\caption{\bf Comparison of Estimated Coefficients and Smooth Terms across Copula Models}
\label{table:copula_compare}
\begin{tabular}{llcccccc}
\toprule
\multicolumn{2}{l}{} &
\multicolumn{2}{c}{\textbf{Gaussian (pp)}} &
\multicolumn{2}{c}{\textbf{Joe (pp)}} &
\multicolumn{2}{c}{\textbf{Clayton 180 (pl)}} \\
\midrule
\multicolumn{8}{l}{\textbf{Treatment Equation}} \\
\midrule
\multicolumn{8}{l}{\textit{Parametric Coefficients}}  \\
\multicolumn{2}{l}{\textbf{Variable}} &\textbf{Estimate} & \textbf{P-value} &\textbf{Estimate} & \textbf{P-value} & \textbf{Estimate} & \textbf{P-value} \\
\midrule
Intercept &  & -0.2315 & 0.394 & -0.2264 & 0.400 & -0.2254 & 0.401 \\
Regional lymph node metastasis &  & 0.6734 & 0.0011 ** & 0.6968 & 0.0011 ** & 0.6940 & 0.0011 ** \\
Margin status (Negative) &  & -1.0920 & 3.07e-05 *** & -1.1505 & 1.09e-05 *** & -1.1456 & 1.14e-05 *** \\
Margin status (Close) &  & -0.8416 & 0.0266 * & -0.7833 & 0.0357 * & -0.7879 & 0.0342 * \\
Perineural invasion (Yes) &  & 0.5780 & 0.0031 ** & 0.5798 & 0.0037 ** & 0.5773 & 0.0038 ** \\
\midrule
\multicolumn{8}{l}{\textit{Smooth Components’ Approximate Significance}} \\
\multicolumn{2}{l}{\textbf{Variable}} &\textbf{edf} & \textbf{P-value} &\textbf{edf} & \textbf{P-value} & \textbf{edf} & \textbf{P-value} \\
\midrule
hsa-miR-203a-3p &  & 2.220 & 0.00029 *** & 2.242 & 0.00018 *** & 2.222 & 0.00018 *** \\
hsa-miR-194-5p &  & 2.799 & 0.01281 * & 2.988 & 0.0069 ** & 2.977 & 0.0074 ** \\
Age at diagnosis &  & 1.000 & 0.03249 * & 1.000 & 0.0283 * & 1.000 & 0.0307 * \\
\midrule
\midrule
\multicolumn{8}{l}{\textbf{Outcome Equation}} \\
\midrule
\multicolumn{8}{l}{\textit{Parametric Coefficients}} \\
\multicolumn{2}{l}{\textbf{Variable}} &\textbf{edf} & \textbf{P-value} &\textbf{edf} & \textbf{P-value} & \textbf{edf} & \textbf{P-value} \\
\midrule
Intercept &  & 0.7067 & 7.49e-08 *** & 0.7323 & 2.66e-09 *** & 1.2365 & 6.39e-09 *** \\
Lymphovascular invasion (Yes) &  & -1.4457 & 1.62e-07 *** & -1.4231 & 1.65e-09 *** & -2.4107 & 1.51e-09 *** \\
\midrule
\multicolumn{8}{l}{\textit{Smooth Components’ Approximate Significance}} \\
\multicolumn{2}{l}{\textbf{Variable}} &\textbf{edf} & \textbf{P-value} &\textbf{edf} & \textbf{P-value} & \textbf{edf} & \textbf{P-value} \\
\midrule
hsa-miR-337-3p &  & 1.000 & 0.00123 ** & 1.000 & 0.00082 *** & 1.000 & 0.00086 *** \\
hsa-miR-145-3p &  & 1.000 & 0.01650 * & 1.305 & 0.0318 * & 1.138 & 0.02086 * \\
hsa-miR-99a-5p &  & 1.000 & 0.00228 ** & 1.177 & 0.0057 ** & 1.095 & 0.00448 ** \\
hsa-miR-146b-3p &  & 1.000 & 0.02392 * & 1.000 & 0.0215 * & 1.000 & 0.02943 * \\
Age at diagnosis &  & 1.000 & 0.00460 ** & 1.000 & 0.0057 ** & 1.000 & 0.00799 ** \\
\bottomrule
\end{tabular}

\vspace{1ex}
\raggedright
\footnotesize
\textit{edf (estimated degrees of freedom) indicates the effective nonlinearity of each smooth term in the model; 
***$p<0.001$, **$p<0.01$, *$p<0.05$. pp, probit--probit margins; pl, probit--logit margins. 
Copula parameter estimates: Gaussian: $\rho = 0.636$ (95\% CI: 0.173--0.866); 
Joe: $\theta = 2.74$ (95\% CI: 1.67--7.24); 
Clayton 180: $\theta = 1.94$ (95\% CI: 0.526--5.10); $n = 215$.
}
\end{table}

As shown in Table \ref{tab:table6}, the overall pattern of associations between covariates and 2-year survival was similar for the one-stage logistic model and the outcome margin of the Gaussian copula model. In both models, LVI was strongly and inversely associated with 2-year survival (p= 1.62e-07), and the estimated effect was somewhat more pronounced in the copula model (estimate -1.45 vs. -1.25 in the one-stage logistic regression). The miRNA expression variables and age at diagnosis showed consistent directions and levels of statistical significance across the two approaches, and all smooth terms in the copula model had edf equal to 1, indicating approximately linear effects that align with the linear terms in the logistic model. Corresponding estimates from the two-stage logistic models are reported in Supplementary Table S12.

\begin{table}[H]
\centering
\caption{Comparison of estimated coefficients and smooth terms between the one-stage logistic model and the Gaussian copula model (probit--probit margins).}
\label{tab:table6}
\begin{tabular}{lcccc}
\toprule
 & \multicolumn{2}{c}{One-stage logistic} & \multicolumn{2}{c}{Gaussian (pp)} \\
\cmidrule(lr){2-3}\cmidrule(lr){4-5}
\multicolumn{5}{l}{\textit{Parametric coefficients}}\\

\textbf{Variable} & \textbf{Estimate} & \textbf{P-value} & \textbf{Estimate} & \textbf{P-value} \\
\midrule
Intercept & 2.08231 & 0.293733 & 0.7067  & 7.49e-08** \\
Lymphovascular invasion (Yes) & -1.24883 & 0.000223*** & -1.4457 & 1.62e-07*** \\
\midrule
\multicolumn{5}{l}{\textit{miRNA expression and age at diagnosis}}\\

\textbf{Variable} & \textbf{Estimate} & \textbf{P-value} & \textbf{edf} & \textbf{P-value} \\
\midrule
hsa-miR-337-3p & -0.66754 & 0.000527*** & 1.000 & 0.00123** \\
hsa-miR-145-3p & 0.61154  & 0.045269* & 1.000 & 0.01650* \\
hsa-miR-99a-5p & 0.41072  & 0.002001** & 1.000 & 0.00228** \\
hsa-miR-146b-3p & -0.38308 & $0.069057$ & 1.000 & 0.02392* \\
Age at diagnosis & -0.03082 & 0.024446* & 1.000 & 0.00460** \\
\bottomrule
\end{tabular}

\vspace{0.5em}
\raggedright
\footnotesize
\textit{One-stage logistic refers to a standard logistic regression of 2-year survival status (alive vs.\ dead) on LVI, miRNA expression, and age.
\textit{Gaussian (pp)} refers to a Gaussian copula model that jointly models LVI and 2-year survival with probit--probit margins. edf (estimated degrees of freedom; Gaussian copula model only) measures the complexity of each spline term edf = 1 $\approx$ linear; edf $>$ 1 indicates nonlinearity).
***$p<0.001$, **$p<0.01$, *$p<0.05$. pp, probit--probit margins; $n=215$.} 
\end{table}

\begin{figure}[H]
    \centering
    \includegraphics[width=1\linewidth]{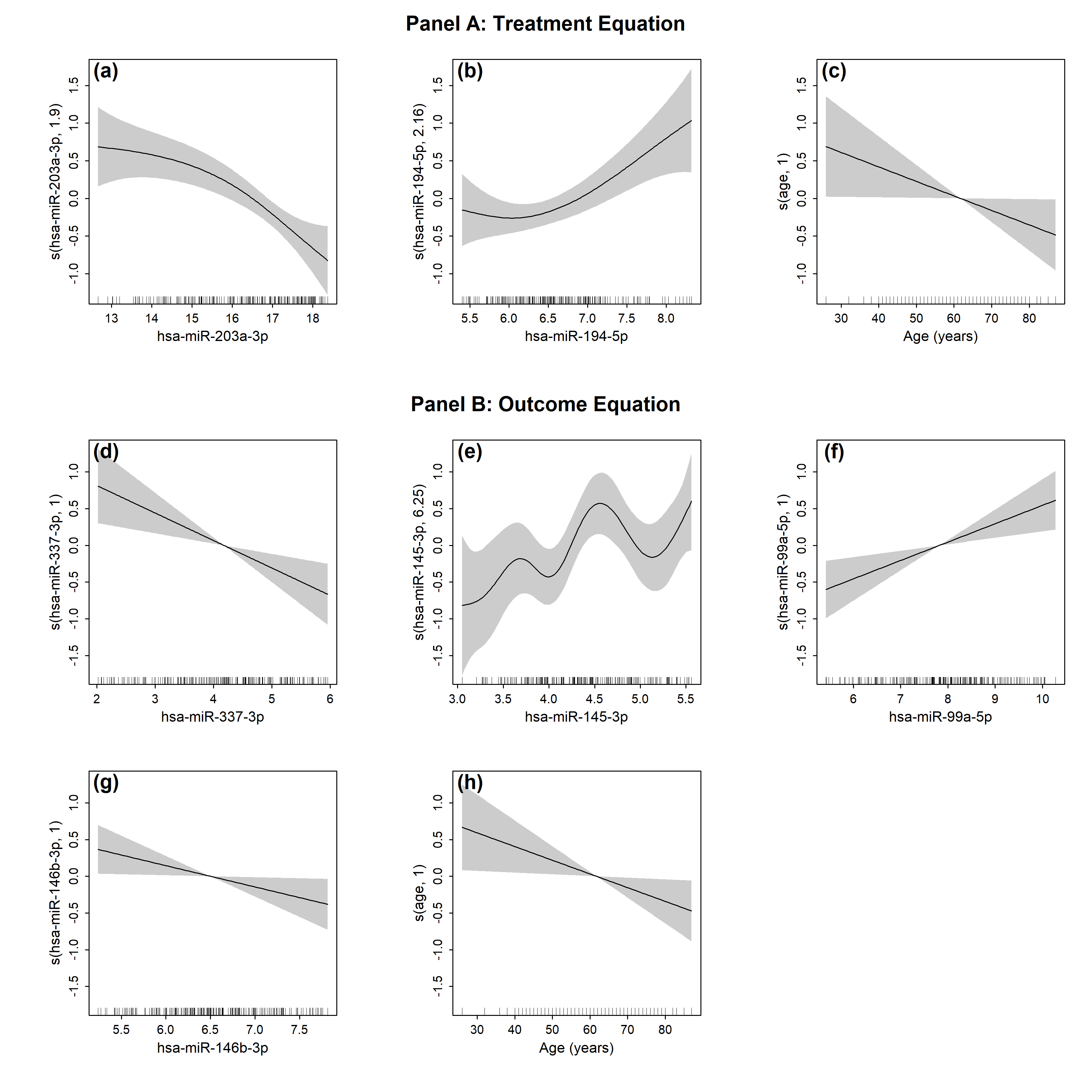}
    \caption{\textit{Estimated smooth effects from the bivariate Gaussian copula model with probit–probit link functions. Panel A shows the treatment equation results, modeling the probability of LVI as a function of selected miRNA expression levels and clinical covariates. Panel B displays the outcome equation results, modeling the probability of 2-year survival. Shaded areas represent 95\% confidence intervals.}}
    \label{fig:normal p-p smooth plot}
\end{figure}

Figure \ref{fig:normal p-p smooth plot} presents the estimated smooth effects from the treatment and outcome equations under the Gaussian copula model, and corresponding results for the Joe and Clayton 180 copula models can be found in Supplementary Figure S1 and S2. In each figure, Panel A shows the smooth components from the treatment equation, modeling the probability of LVI as a function of selected miRNA expression levels and age at diagnosis. Panel B displays the smooth components from the outcome equation, modeling the probability of 2-year survival. Across all models, several continuous covariates demonstrated statistically significant nonlinear associations in at least one equation, with consistent patterns observed for key variables.

In the treatment equation (Figure \ref{fig:normal p-p smooth plot}A), hsa-miR-203a-3p and hsa-miR-194-5p showed clear nonlinear associations with the probability of LVI. The smooth function of hsa-miR-203a-3p demonstrated a downward trend, indicating that higher expression was associated with reduced likelihood of LVI, whereas hsa-miR-194-5p exhibited an upward slope, particularly pronounced at higher expression levels. Age at diagnosis displayed a linear negative effect, with older age consistently linked to a lower probability of LVI.

In the outcome equation (Figure \ref{fig:normal p-p smooth plot}B), hsa-miR-337-3p, hsa-miR-145-3p, hsa-miR-99a-5p, hsa-miR-146b-3p, and age at diagnosis all showed measurable effects on the probability of 2-year survival. Most smooth functions were approximately monotonic, with hsa-miR-337-3p and hsa-miR-146b-3p both displaying negative associations with 2-year survival probability, whereas hsa-miR-99a-5p showed a positive relationship, suggesting protective effects of higher hsa-miR-99a-5p expression. Notably, hsa-miR-145-3p exhibited a complex, undulating pattern, with alternating increases and decreases across the expression range, accompanied by wider CIs that reflected greater uncertainty. The smooth term for age exhibited a consistent decline, reinforcing that older patients had a reduced probability of surviving for 2 years.

\section*{Discussion}

In this study, we used a semiparametric recursive copula model to estimate the effect of LVI on 2-year survival in patients with HNSC, adjusting for miRNA expression and clinical covariates. Our findings indicate that patients with LVI have substantially lower 2-year survival probabilities than patients without LVI, with an SATE of approximately -47\%. At the molecular level, expression levels of miRNAs, such as hsa-miR-337-3p, hsa-miR-145-3p, and hsa-miR-99a-5p, were significantly associated with survival, supporting prior evidence that miRNAs regulate tumor progression and metastasis by modulating key oncogenic signaling pathways \cite{grzywa2020regulators}.

Compared with traditional one- and two-stage regression methods, the copula model can flexibly handle potential endogeneity between LVI and survival \cite{dettoni2023effect}. The results from our simulation studies show that the copula model was more robust and was associated with more negligible estimation bias than other models under different levels of endogeneity and censoring, with particularly enhanced performance under conditions of medium–high censoring and strong endogeneity. In addition, this study used semiparametric models (such as nonlinear spline functions) to model continuous variables, effectively capturing the nonlinear relationship between covariates, such as miRNA expression, age, and survival, better representing the actual clinical situation than other models.

The estimated Kendall’s $\tau$ values are uniformly positive across copula models. Prior studies have interpreted such positive dependence as evidence of a direct positive association between the treatment, $Y_1$ (LVI status), and the outcome, $Y_2$ (two-year survival). However, in our setting, this interpretation is implausible because an increased risk of LVI is unlikely to causally improve survival. Instead, we interpret the copula dependence as capturing correlations among unobserved disturbances in the two equations (Equations~\ref{eq:y1star} and \ref{eq:y2star}), conditional on observed covariates. In recursive or joint models, the dependence parameter summarizes the associations among unobserved errors in the treatment and outcome equations [e.g., $\operatorname{Cov}(\varepsilon_1,\varepsilon_2)=\rho$ in a bivariate probit]. Under a copula specification, Kendall’s $\tau$ is a monotone transform of this cross-equation error-term dependence. Therefore, a positive $\tau$ is more likely to indicate positively correlated unobservable variables across equations than to indicate a marginal causal benefit of the treatment \cite{radice2016copula}.

From a clinical and biological perspective, our results demonstrate that LVI is significantly associated with regional lymph node metastasis, resection margin status, and perineural infiltration, consistent with existing studies and supporting LVI as a marker of an aggressive tumor phenotype with poorer survival. Importantly, these co-occurring adverse clinicopathologic features suggest that LVI may not behave as an exogenous exposure: latent tumor aggressiveness and other unmeasured biological processes can simultaneously increase the probability of observing LVI and worsen survival, thereby inducing residual dependence (endogeneity) between the LVI and survival processes. We further explored potential molecular mechanisms by examining miRNAs associated with survival and LVI to provide molecular context for this potential endogeneity. hsa-miR-203a-3p is predictive of LVI risk in HNSC and may modulate fatty acid oxidation pathways, lymphangiogenesis-related genes, or Hippo signaling to promote invasive and metastatic behaviors \cite{karmakar2021identification}. hsa-miR-194-5p overexpression can regulate chromodomain helicase DNA-binding protein 4 and activate the PI3K/AKT signaling pathway, enhancing tumor cell resistance to apoptosis and correlating with worse outcomes in oral squamous cell carcinoma \cite{li2023bioinformatics}. hsa-miR-145-3p functions as a tumor suppressor in HNSC cells by targeting myosin-Ib to inhibit cell proliferation and migration \cite{yamada2018passenger}, and hsa-miR-99a-5p is often downregulated in HNSC, with restoration suppressing proliferation and migration \cite{sun2021microrna}. hsa-miR-337-3p has appeared in prognostic miRNA signatures in HNSC and may be linked to invasion capacity or lymphangiogenesis-related targets \cite{wu2021mirna}. The role of hsa-miR-146b-3p in HNSC is less clear, with evidence suggesting context-dependent involvement in immune regulation or migration-related pathways \cite{lan2023ptpn12}. Overall, these miRNAs point to upstream biological programs related to proliferation, resistance to apoptosis, invasion and migration, immune regulation, and lymphatic or vascular remodeling that plausibly drive both LVI occurrence and survival outcomes, supporting the biological plausibility of endogenous LVI in HNSC. The nonlinear associations between miRNA expression levels and clinical outcomes highlight the complexity of cancer biology \cite{mukherji2011micrornas}.

Despite the strengths of the present study, certain limitations must be acknowledged. First, the data come from the TCGA database, and some patients have incomplete follow-up or are missing key clinical data, which may introduce selection or information bias. Second, adapting survival data for GJRM via conversion into binary variables may lose some detailed information, as the transformation removes temporal variation in event and censoring times \cite{salika2022implications}. This simplification reduces efficiency and limits interpretation to a fixed-time survival snapshot rather than the full time-to-event process. Third, some assumptions of the copula model, such as the selection of specific copula functions and marginal distributions, may impact the results, as different copula families capture distinct dependence structures, and misspecification may bias the estimated association or treatment effects. Likewise, incorrect marginal choices can distort parameter estimates and inference, leading to potentially different substantive conclusions.

In conclusion, our study identified a significant negative impact of LVI on 2-year survival probability among HNSC patients by applying a semiparametric recursive copula model, effectively addressing the issue of endogeneity between LVI and survival outcomes. Our findings also identified miRNAs that may play complex regulatory roles in HNSC, including hsa-miR-203a-3p, hsa-miR-194-5p, hsa-miR-99a-5p, and hsa-miR-337-3p, providing potential molecular insights that can be used to inform personalized treatment strategies and risk stratification. Although this study highlights the advantages of the copula approach over traditional regression methods, further research is necessary to verify the generalizability of different copula functions and explore the biological mechanisms underlying these associations at greater depth.

\section*{Methods}

\subsection*{Data Source and Study Population}

Data for this study were obtained from the TCGA-HNSC project. Of 604 patients with survival information, 506 had corresponding miRNA expression data. Patients were included only if they had clearly documented LVI status and complete key clinical variables, including tumor stage, grade, and survival outcome. After data cleaning and preprocessing, including the removal of cases with missing values, a total of 215 patients were retained for analysis. miRNA expression data were sourced from the University of California, Santa Cruz Xena database (\textit{https://xenabrowser.net/}) and processed into level 3 data representing normalized miRNA expression estimates (in reads per million [RPM]), derived from aligned, quality-controlled RNA sequencing data. For each mature miRNA, isoform-level expression values were aggregated and log-transformed as $log_2\text{(total RPM + 1)}$ to reduce skewness and enhance the stability of the expression data for statistical analysis.

\subsection*{Selection of miRNA screening methods}
Understanding the causal effect of LVI on cancer survival is complicated by both potential endogeneity and the high dimensionality of genetic data. We preprocessed genetic data to reduce this dimensionality by screening miRNAs associated with survival outcomes \cite{fan2008sure,wang2022feature}. 

The miRNA screening occurred in two stages. First, we used a series of statistical methods to preliminarily screen miRNAs related to patient survival. By using multiple methods, we are able to systematically evaluate the consistency of results across methods and explore differences among theoretical models and statistical models. Previous studies have found that the selection of miRNA screening methods is crucial for ensuring a deep understanding of the molecular characteristics underlying cancer progression mechanisms and patient survival prognosis (46). During the initial screening stage, we applied semiparametric and parametric methods, including partial likelihood (PL), sure independence screening (SIS), and feature aberration at survival times; distance correlation-based methods, including robust censored distance correlation screening (RCDCS) and composite RCDCS; and nonparametric ranking-based approaches, including the concordance index (CINDEX) and inverse probability of censoring weighted Kendall’s tau (IPCW-Tau). We also applied an extended version of IPCW-Tau that integrates gene–gene network structures through a graph-based transformation, enabling the screening procedure to account for underlying biological dependencies and enhance the relevance of selected biomarkers  \cite{fan2010high,saldana2018sis,gorst2012coordinate,chen2018robust,harrell1996multivariable,song2014censored}. The candidate miRNAs identified across all of these methods were highly consistent, with a high degree of overlap.

In the second stage, we focused on identifying miRNAs correlated with LVI status, for which we applied the SIS method, which is suitable for handling binary outcomes \cite{saldana2018sis}. To verify the stability and reliability of our screening results, we referenced and reproduced the random forest algorithm used to identify miRNAs related to LVI in existing literature \cite{karmakar2021identification}. Our identified miRNAs were consistent with those reported in the literature, including some miRNAs previously validated as being involved in metastatic mechanisms involved in LVI \cite{majumder2020fragile,holt2021integrative}. Similar to the results of the first stage, the miRNAs identified as being related to LVI by each method displayed substantial overlap, with only a few miRNAs uniquely identified by individual methods. 

Combining the results from both stages, we ultimately identified 24 candidate miRNAs for further analysis. All detailed screening outcomes and method-specific lists are provided in Supplementary Material, Sections S1 and S2.

\subsection*{Copula Model Formulation}

To assess the causal relationship between LVI and survival status while addressing potential endogeneity from unmeasured factors, we applied recursive copula models. Recursive copula additive models extend the simultaneous estimation framework, using copula construction to ensure valid likelihood-based inference (39,55), providing a likelihood-based framework for parameter estimation and causal effect evaluation. When one variable affects another, the recursive structure is advantageous, allowing the model to capture this direct effect while accounting for endogeneity and ensuring that the dependency between the endogenous variable and the outcome is modeled correctly. Identification of the recursive copula model with two binary margins has been extensively discussed \cite{marra2020estimating, radice2016copula}.

We aim to estimate the effect of LVI status (denoted $Y_{i1}$) on survival status $Y_{i2}$, which are both binary variables. The model assumes that the observed variables, $Y_{i1}$ and $Y_{i2}$, are dichotomized versions of continuous latent variables, as follows.

We consider two binary observed outcomes $Y_{ij}$ ($j = 1,2$), which are assumed to arise from dichotomizing latent continuous variables $\widetilde{Y}_{ij}$:

\begin{eqnarray}
Y_{ij} &=& 
\begin{cases}
1, & \text{if } \widetilde{Y}_{ij} > 0, \\ 
0, & \text{otherwise},
\end{cases}
\label{eq:dichot}
\end{eqnarray}

for $i=1,\ldots,n$ and $j=1,2$. The first latent variable, $\widetilde{Y}_{i1}$, represents the unobserved propensity toward LVI (treatment), whereas  $\widetilde{Y}_{i2}$ reflects the latent survival tendency (outcome). We specify the following recursive additive model for these latent responses:

\begin{eqnarray}
\widetilde{Y}_{i1} &=& \eta_1(\mathbf{X}_{i1}) + \epsilon_{i1}, 
\label{eq:y1star} \\
\widetilde{Y}_{i2} &=& \gamma Y_{i1} + \eta_2(\mathbf{X}_{i2}) + \epsilon_{i2}, 
\label{eq:y2star}
\end{eqnarray}
where $\eta_j(\mathbf{X}_{ij})$ denotes the additive effects of covariates on the latent variable $\widetilde{Y}_{ij}$ for $j=1,2$, and $Y_{i1}$ is the observed binary treatment status. Each additive term $\eta_j(\mathbf{X}_{ij})$ is modeled as:
\begin{eqnarray}
\eta_j(\mathbf{X}_{ij}) &=& \alpha_{j,0} + \sum_{k=1}^{K_j} h_{jk}(X_{ijk}),
\label{eq:addterm}
\end{eqnarray}
where $\alpha_{j,0}$ is the intercept, $K_j$ is the number of covariates in the $j$-th equation, and $h_{jk}(X_{ijk})$ represents the potentially nonlinear effect of the $k$-th covariate in equation $j$.

To flexibly capture nonlinear relationships, we approximate each smooth function $h_{jk}(X_{ijk})$ using a basis expansion:
\begin{eqnarray}
h_{jk}(X_{ijk}) &=& \sum_{l=1}^{\nu_{jk}} \beta_{jkl} B_{jkl}(X_{ijk}),
\label{eq:basis}
\end{eqnarray}
where $B_{jkl}(\cdot)$ is the $l$-th basis function associated with the $k$-th covariate in the $j$-th equation, and $\nu_{jk}$ is the number of basis functions used. The covariates, $\mathbf{X}_{i1}$ and $\mathbf{X}_{i2}$, are potentially overlapping subsets of the full covariate vector $\mathbf{X}_i$ (i.e., $\mathbf{X}_i = \mathbf{X}_{i1} \cup \mathbf{X}_{i2}$). We assume that the error terms, $\boldsymbol{\varepsilon}_i = (\varepsilon_{i1}, \varepsilon_{i2})^\top$, are independent and identically distributed across individuals, with zero mean and finite variance. The dependence between the two equations is captured through the correlation $\rho = \text{Corr}(\epsilon_{i1}, \epsilon_{i2})$, which is modeled using a parametric copula function.

Copula modeling separates the specification of the marginals from the dependence structure, with the copula function $C$ linking them into a joint distribution \cite{dettoni2023effect}. Here, we opt for a joint parametric copula model. A copula model specifies a joint cumulative distribution function for two or more variables. Based on the latent variable formulation in Equations~\ref{eq:dichot}–\ref{eq:y2star}, the marginal success probability for each binary outcome is given by

\begin{eqnarray}
P(Y_{ij}=1) &=& 1 - F_j\big(-\eta_j(\mathbf{X}_{ij})\big),
\qquad j=1,2
\label{eq:marg}
\end{eqnarray}
where $F_j(\cdot)$ denotes the cumulative distribution function (CDF). The marginal CDFs are modeled using logit, probit, or cloglog links, with covariates entering through the predictor $\eta_j\left(\mathbf{X}_{i j}\right)$. To construct the joint distribution, we employ a copula function $C_\theta: (0,1)\times(0,1)\to(0,1)$. Accordingly, the joint probability of $(Y_{i1},Y_{i2})$ is
\begin{eqnarray}
P(Y_{i1}=1, Y_{i2}=1) &=& C_\theta\!\left(P(Y_{i1}=1),\; P(Y_{i2}=1)\right),
\label{eq:copula}
\end{eqnarray}
where $\theta$ denotes a copula model parameter. The dependence between $Y_{i1}$ and $Y_{i2}$, after accounting for the covariate effects on the marginals, is described by the association parameter $\theta$.

In the GJRM framework, various copulas are implemented, each associated with a parameter $\theta$ that governs the dependence structure between the variables. Supplementary Table S21 provides the range of $\theta$ values and outlines the process for calculating Kendall’s $\tau$.

The primary estimand of interest is the SATE, which quantifies the expected change in survival probability attributable to LVI. Following Radice \cite{marra2020estimating}, we define the binary outcome variables $Y_{i1}$ and $Y_{i2}$ as taking values in ${0,1}$, leading to four possible joint configurations:$\kappa(Y_{i1}=1, Y_{i2}=1) = C_\theta(p_{i1}, p_{i2})$, $\kappa(Y_{i1}=1, Y_{i2}=0) = p_{i1} - C_\theta(p_{i1}, p_{i2})$,$\kappa(Y_{i1}=0, Y_{i2}=1) = p_{i2} - C_\theta(p_{i1}, p_{i2})$, and 
$\kappa(Y_{i1}=0, Y_{i2}=0) = 1 - p_{i1} - p_{i2} + C_\theta(p_{i1}, p_{i2})$.
The corresponding log-likelihood function for the copula model can then be written as:
\begin{align}
\mathcal{L}(\alpha,\beta,\theta)
= \sum_{i=1}^n \Big\{ 
& Y_{i1}Y_{i2}\log[\kappa(1,1)] + Y_{i1}(1-Y_{i2})\log[\kappa(1,0)] \notag \\
&+ (1-Y_{i1})Y_{i2}\log[\kappa(0,1)] + (1-Y_{i1})(1-Y_{i2})\log[\kappa(0,0)]
\Big\} \label{eq:likelihood}
\end{align}
The effect of a treatment on the probability of $Y_{i1}$ affecting $Y_{i2}$ is usually of primary interest. The analysis aims to study how an endogenous variable changes the expected outcome. Thus, the treatment effect is measured by the difference between the expected outcomes of the treated and untreated groups. The SATE function can be defined as
\begin{eqnarray}
\operatorname{SATE}(\alpha, \beta, \theta, X)=\frac{1}{n} \sum_{i=1}^n\left(P\left(Y_{i 2}=1 \mid Y_{i 1}=1, \mathbf{X}_i ; \alpha, \beta, \theta\right)-P\left(Y_{i 2}=1 \mid Y_{i 1}=0, \mathbf{X}_i ; \alpha, \beta, \theta\right)\right)
\label{eq:sate}
\end{eqnarray}
We evaluate the performance of the proposed copula model under varying degrees of endogeneity and censoring through simulation studies, which are presented in the next section.

\subsection*{Simulation Design}

\subsubsection*{Simulation Study 1: variable-selection performance}

The goal of this simulation study was to examine the ability of copula-based joint models to identify truly relevant predictors under two settings: (1) when the model is correctly specified and (2) when the model is misspecified. The simulation design was chosen to broadly resemble the complexity of the case study while remaining fully synthetic.

For each subject $i=1, \ldots, n$, we generated a 10-dimensional covariate vector, $\mathbf{X}_i \sim N(0, \Sigma)$, where $\Sigma$ had unit variances and pairwise correlations of 0.5. To introduce nonlinearity, one covariate was transformed using $f_1(x)=\cos (2 \pi x)+\sin (\pi x)$. The binary treatment indicator $Y_{1i}$ was modeled with a probit link,
\[
Y_{1 i}=\mathbf{1}\{\eta_{1 i}+\varepsilon_{1 i}>0\}, \quad \eta_{1 i}=\mathbf{b}_i^\top \beta, \quad \varepsilon_{1 i}\sim N(0,1).
\]
For the fully specified model, the coefficient vector was set to $\beta_{\text{full}}=(0.9, 1, 1.4, 0.7, -1.08, 0.6)^\top$, whereas in the reduced, misspecified model, several coefficients were set to zero, yielding $\beta_{\text{reduced}}=(0, 1, 1.4, 0, -1.08, 0)^\top$. Survival times were generated from a log-normal accelerated failure time (AFT) model,
\[
\log(t_i)=Z_i^\top\gamma + \varepsilon_{2i}, \quad \varepsilon_{2i}\sim N(0,1),
\]
with covariate vector $Z=(Y_1, x_3, x_7, x_8, x_9, x_{10})^{T}$. 

In the fully specified model, the coefficient vector was $\gamma_{\text{full}}=(-1.39,-1.31,-1.40,1.12,1.60,-1.23)^{T}$; in the reduced, misspecified model, coefficients for $x_7, x_9$, and $x_{10}$ were set to zero. Dependence between the treatment and survival error terms was induced through a Gaussian copula, with correlation $\rho=0.5$. Right censoring was applied using independent uniform censoring times to achieve the desired censoring rates.  Observed survival outcomes were then dichotomized at a prespecified cutoff, $t_i$, producing the binary survival status $Y_{2i}$.

Each simulation scenario consisted of $m=200$ replicates with $n=200$ per replicate. Performance was evaluated in terms of the ability to recover the true active set of predictors, along with secondary measures such as SATE, Kendall’s $\tau$, and confidence intervals. Convergence diagnostics were monitored throughout.

\subsubsection*{Simulation Study 2: treatment-effect estimation accuracy}

The second simulation study aimed to evaluate the accuracy and reliability of treatment effect estimation under different modeling strategies. Specifically, we compared the one-stage Cox proportional hazards (PH) model (Cox PH), 2SPS, and the copula-based joint regression model with respect to their ability to recover the coefficient of $Y_1$, which was fixed at -1.39 in the data-generating process. We considered three sample sizes (n=200,500,1000), three survival cutoffs (0.25, 0.5, 0.75), and censoring rates ranging from 0\% to 60\%. Each unique setting was replicated 500 times. Although the survival times were primarily generated from an AFT model, we evaluated both Cox PH and AFT estimators on these data and found broadly consistent results. As an additional robustness check, we generated survival outcomes under a Cox PH specification and observed similar performance across estimators.
For the one-stage method, a Cox model was fit with covariates $X_i=(Y_{i1}, X_{i3}, X_{i7}, X_{i8}, X_{i9}, X_{i10})^{T}$, and the coefficient of Y  was taken as the target parameter. The first stage of the two-stage approach employed a logistic regression for $Y_1$  to obtain fitted values, $\hat{Y_1}$, which were then included as predictors in the second-stage regression for the survival outcome. Although two-stage residual inclusion (2SRI) is sometimes used for binary outcomes, both 2SLS and 2SRI rely on strong instruments; in the absence of such instruments, their performance is essentially indistinguishable \cite{terza2008two}. Finally, the copula-based model was estimated by fitting a recursive copula system for ($Y_1$, $Y_2$) at each cutoff and extracting the outcome equation coefficient on $Y_1$. Performance across methods was assessed in terms of bias, estimation accuracy, and stability under varying sample sizes and levels of censoring.

\section*{Acknowledgements (not compulsory)}

This research was supported in part by NIH grant R01DK136994 (PI: Roger S. Zoh, Indiana University). \\

Additional support was provided by NIH grant R01DK132385 (PI: Carmen D. Tekwe, Indiana University).

\section*{Data availability}
All statistical analyses were performed using R version 4.3.1. The analysis employed the following key packages:
\begin{itemize}
\item \texttt{GJRM} version 0.2-6.5 for copula-based joint regression modeling
\item \texttt{randomForest} version 4.7-1.1 for machine learning approaches and feature importance analysis
\end{itemize}
The complete implementation code, simulation application scripts, and processed datasets are publicly available in our GitHub repository: \url{https://github.com/young-moredcai/LVI_2yrSurvival_Copula_TCGA_HNSC.git}. 

\section*{Additional information}

\textbf{Accession codes:} Not applicable. \\
\textbf{Competing interests:} The authors declare no competing interests.

\bibliography{sample} 

\end{document}